\documentclass{aa}
\usepackage[dvips]{graphicx}
\begin{document}

\title{The mass of the neutron star in SMC X-1}

\titlerunning{The mass of the neutron star in SMC X-1}

\author{A.K.F. Val Baker \and  A.J. Norton \and 
H. Quaintrell}

\authorrunning{A.K.F. Val Baker et al}

\offprints{A.J. Norton, a.j.norton@open.ac.uk}

\institute{Department of Physics and Astronomy, The Open University, 
        Walton Hall, Milton Keynes MK7 6AA, U.K.}

\date{Accepted 01/07/2005
      Received 16/03/2005}

\abstract{We present new optical spectroscopy of the eclipsing 
binary pulsar Sk~160/SMC~X-1. From the He~I absorption lines, taking
heating corrections into account, we determine the radial velocity 
semi-amplitude of Sk~160 to be $21.8 \pm 1.8$~km~s$^{-1}$. 
Assuming Sk~160 fills its Roche-lobe, the inclination angle of the 
system is $i=65.3^{\circ} \pm 1.3^{\circ}$ and in this case we obtain 
upper limits for the mass of the neutron star as $M_{\rm x}= 1.21 \pm 
0.10$~M$_{\odot}$ and for Sk 160 as $M_{\rm o}= 16.6 \pm 0.4$~M$_{\odot}$. 
However if we assume that the inclination angle is $i=90^{\circ}$, then the 
ratio of the radius of Sk~160 to the radius of its Roche-lobe is $\beta = 
0.79 \pm 0.02$, and the lower limits for the masses of the two stars are 
$M_{\rm x}= 0.91 \pm 0.08$~M$_{\odot}$ and $M_{\rm o}= 12.5 \pm 
0.1$~M$_{\odot}$. We also show that the He~II 4686\AA \ emission line
tracks the motion of the neutron star, but with a radial velocity amplitude 
somewhat less than that of the neutron star itself. We suggest that this 
emission may arise from a hotspot where material accreting via Roche lobe 
overflow impacts the outer edge of an accretion disc.
 
\keywords{
 binaries: close -- stars: neutron -- 
 stars: individual: SMC X-1 -- 
 stars: individual: Sk 160 --
 stars: fundamental parameters
 }}

\maketitle

\section{Introduction}
 
Eclipsing X-ray pulsars offer a means of directly measuring neutron star 
masses. However, only 7 such systems are currently known and the neutron 
star masses in each case are not determined to high accuracy.  If the 
situation can be improved, the equation of state for nuclear matter may be 
constrained, so testing theories that describe it. To this end that we have 
carried out the present study of SMC X-1. The determination of the system 
masses follows from measuring the semi-amplitude of the neutron star's radial 
velocity (RV) curve ($K_{\rm x}$, from X-ray pulse timing delays), the X-ray 
eclipse half angle ($\theta_{\rm e}$), and the semi-amplitude of the 
companion star's RV curve ($K_{\rm o}$, from absorption line optical 
spectroscopy); for details see e.g. Ash et al. (1999) and Quaintrell et al. 
(2003). It is important to note that the relevant equations can only be solved 
by assuming a value for either the system inclination angle ($i$) or the ratio 
of the radius of the optical companion star to that of its Roche lobe 
($\beta$), see Equation (4) of Quaintrell et al. (2003). Previous studies of 
SMC X-1 have generally assumed the companion star is Roche lobe-filling 
(i.e. $\beta = 1$) and so obtained masses that are only relevant in this 
limit; we return to this later.

The optical counterpart to SMC X-1 is the B0~I supergiant, Sk~160 (Webster et 
al. 1972; Liller 1972). The X-ray source has a pulse period of 0.72~s and
exhibits an eclipse duration of $0.610 \pm 0.019$~d (Primini et al. 1976) 
in the 3.892~d orbit, corresponding to $\theta_{\rm e} = 28.2^{\circ} 
\pm 0.9^{\circ}$. Timing studies of the X-ray pulsations (Levine et al. 1993) 
give $(a_x \sin i) = 53.4876 \pm 0.0004$~lt~sec for the projected semi-major 
axis and indicate a circular orbit with $e < 0.00004$. The corresponding RV 
amplitude is $K_{\rm x} = 299.607 \pm 0.002$~km~s$^{-1}$. The X-ray emission 
from SMC X-1 has also been found to exhibit a long quasi-stable 
super-orbital period of 50--60 days, believed to be a result of obscuration 
of the neutron star by a precessing accretion disk (e.g. Wojdowski et al. 
1998). The mass transfer in SMC X-1 probably has significant contributions 
from Roche-lobe overflow (Khruzina \& Cherepashchuk 1983; van Paradijs \& 
Kuiper 1984), as the stellar winds observed in Sk 160 are not strong enough 
to power the accretion (Hammerschlag-Hensberge, Kallman \& Howarth 1984). 

Previous attempts to derive the orbital parameters of SMC X-1 have been made 
by Primini et al. (1976), Hutchings et al. (1977), Reynolds et al. (1993) and  
van der Meer et al. (2005). Reynolds et al. (1993) were the first to account 
for heating of the donor star by the X-ray flux from the neutron star. This 
heating has the effect of significantly altering the observed RV 
amplitude and so distorting the inferred neutron star mass. However van 
Kerkwijk et al. (1995) pointed out the 
uncertainties introduced in this approach by not allowing for the presence of 
an accretion disk, whose shadow on the face of the giant star may reduce the 
effect of X-ray heating. The most recent analysis by van der Meer et al. 
(2005), like several earlier investigations, found a low value for the 
neutron star mass, $M_{\rm x} = 1.05 \pm 0.09$~M$_{\odot}$, but they too did 
not account for any heating correction in their analysis. It should
also be noted that each of the previous mass determinations implicitly 
assumed the companion star to fill its Roche lobe in order to solve for the 
system parameters, so the masses are in effect upper limits in each case.

\section{Observations}

Our observations were obtained from $30^{\rm th}$ August -- 
$18^{\rm th}$ September 2000 using the 1.9 metre Radcliff telescope at the 
Sutherland Observatory. The grating spectrograph was used with 
a reciprocal dispersion of 0.5~\AA / pixel, 
spanning the wavelength range 4300 -- 5100 \AA. Over the course of 3 
weeks (1 week on, 1 week off, 1 week on) we obtained 56 usable spectra 
of Sk~160 on 9 nights, mostly during the first week of observations (see 
Table 2). We note that these observations just preceded the coordinated 
HST/Chandra campaign on SMC X-1 reported by Vrtilek et al. (2001), and 
occurred during a low state of the $\sim 55$~d super-orbital cycle, as 
indicated by the {\em RXTE} ASM lightcurve.

We also observed the RV standard star HD6655, an F8~V star with 
an accurately known radial velocity of $+15.5$~km~s$^{-1}$, on each night. In
addition, on the last night, we observed a template star of similar spectral 
type to Sk~160, for cross-correlation with our target spectra. This was 
HR1174, a B3~V star, which was also used as the cross-correlation template by 
Reynolds et al. (1993).

\section{Data Reduction}

All spectra were reduced using standard {\sc iraf}\footnote{{\sc iraf} is 
distributed by the National Optical Astronomy Observatory, which is operated
by the Association of Universities for Research in Astronomy, Inc., under 
cooperative agreement with the National Science Foundation.} routines; Figure
1 shows a median, continuum normalised spectrum of Sk~160. Note 
that the apparent double peak in the He~II 
4686\AA \ emission line is the result of sampling this line mostly at the 
quadrature phases of the system. The median spectrum therefore shows two 
peaks separated by $\approx 8$\AA \ corresponding to a $\approx 
500$~km~s$^{-1}$ velocity difference.
In order to check the stability of the observations from night to night, we 
cross-correlated the individual spectra of the RV standard star HD6655 against 
a single spectrum of this object from the middle of the run. These were all 
consistent with zero shift from night to night. 

Having confirmed the stability of the system, each individual spectrum of 
Sk~160 was cross-correlated against the median spectrum of the template
star, HR1174. Only regions between 4370--4500\AA, 4700--4735\AA \ and 
4900--5060\AA \ were used, spanning several He~I absorption lines. 
These regions were selected to exclude the Balmer lines which were found to 
show large, random changes from one spectrum to the next, and no clear trend 
in their RVs. It is well known that Balmer lines in high mass stars may be 
contaminated by emission from the star's wind and so may not accurately 
reflect the orbital RV in a binary system. The extent of the emission 
contamination depends on the strength of the stellar wind and is reduced 
for higher level Balmer transitions. Final heliocentric RVs corresponding to 
each spectrum of Sk~160 were calculated from the cross-correlation results 
by applying the heliocentric velocity corrections and then offsetting the 
result by the RV of HR1174,
measured by fitting Gaussians to the He~I lines 
in its spectrum as $+16.7$~km~s$^{-1}$. 
These final RVs are listed 
in Table 2 along with their $1 \sigma$ uncertainties, and plotted in Figure 2.

Orbital phases corresponding to each spectrum were calculated using 
the ephemeris from Wodjdowski et al. (1998), 
which gives the centre time of the $N$th eclipse as $t_N / {\rm MJED} 
= 42836.18278(20) + 3.89229090(43)N - (6.953(28) \times 10^{-8})N^2$.
Numbers in brackets indicate the uncertainties in the last decimal places in
each case. The orbital period at the time of our observations, $\sim 2300$ 
periods after the reference time of this ephemeris, is $P=3.891971(1)$d.
Also included in Figure 2 are the RV measurements of Reynolds et al. (1993), 
with phases calculated according to the revised ephemeris above.

\section{Data analysis}

\subsection{System parameters from the raw He~I RV curve}

The data shown in Figure 2 were fitted with a sinusoid, allowing the mean 
level and amplitude as free parameters. 
The semi-amplitude of the RV curve is $K_{\rm o} = 18.0 \pm 1.8$~km~s$^{-1}$ 
and the systemic velocity is $\gamma  = 174.1 \pm 1.5$~km~s$^{-1}$; these 
values are listed in Table 1.

\begin{table*}
\caption{System parameters for Sk~160/SMC X-1. The two values for 
$K_{\rm o}$ are those resulting from fitting the HeI absorption line RV 
curve without and with heating corrections respectively. The 4 
columns for the inferred parameters are the limiting values assuming
$\beta = 1$ and $i = 90^{\circ}$ for each of the values of 
$K_{\rm o}$ as discussed in the text.}
\begin{tabular}{lllll} \hline
Parameter & \multicolumn{2}{c}{Fit to raw RV curve} & \multicolumn{2}{c}{Fit to corrected RV curve}    \\ \hline
$\gamma$ / km~s$^{-1}$          & \multicolumn{2}{c}{$174.1 \pm 1.5$} & \multicolumn{2}{c}{$173.8 \pm 1.5$}    \\
$K_{\rm o}$ / km s$^{-1}$       & \multicolumn{2}{c}{ $18.0 \pm 1.8$} & \multicolumn{2}{c}{ $21.8 \pm 1.8$}    \\
$q$                             & \multicolumn{2}{c}{$0.060 \pm 0.006$} & \multicolumn{2}{c}{$0.073 \pm 0.006$}\\
                                & Roche lobe-filling & edge-on          & Roche lobe-filling & edge-on \\
                                &(upper mass limits)&(lower mass limits)&(upper mass limits)&(lower mass limits) \\
$\beta$                         & $1.00$          & $0.77 \pm 0.02$     & $1.00$          & $0.79 \pm 0.02$      \\
$i$ / deg                       & $64.0 \pm 1.3$  & $90.0$              & $65.3 \pm 1.3$  & $90.0$               \\
$M_{\rm x}$ / M$_{\odot}$       & $1.01 \pm 0.10$ & $0.73 \pm 0.08$     & $1.21 \pm 0.10$ & $0.91 \pm 0.08$      \\
$M_{\rm o}$ / M$_{\odot}$       & $16.8 \pm 0.5$  & $12.1 \pm 0.2$      & $16.6 \pm 0.4$  & $12.5 \pm 0.1$       \\
$a$ / R$_{\odot}$               & $27.4 \pm 0.3$  & $24.6 \pm 0.2$      & $27.3 \pm 0.3$  & $24.7 \pm 0.1$       \\
$R_{\rm L}$  / R$_{\odot}$      & $16.7 \pm 0.3$  & $14.9 \pm 0.1$      & $16.3 \pm 0.2$  & $14.8 \pm 0.1$       \\
$R_{\rm o}$ / R$_{\odot}$       & $16.7 \pm 0.3$  & $11.5 \pm 0.3$      & $16.3 \pm 0.2$  & $11.7 \pm 0.3$       \\ \hline
\end{tabular}\\
\end{table*}
 
The masses of the stars were calculated following the procedure of Ash et al.
(1999) and Quaintrell et al. (2003), using the Monte Carlo method for 
uncertainty determination described therein. As SMC X-1 has a circular orbit, 
the Roche lobe filling factor 
($\beta = R_{\rm o} / R_{\rm L}$) will not vary, but in the absence of an 
exact value for the radius of the companion star's Roche lobe we cannot 
determine $\beta$ uniquely. Since mass transfer in 
SMC X-1 has significant contributions from Roche lobe overflow, but Sk~160 is 
unlikely to be overfilling its Roche lobe, we can however set an upper limit 
of $\beta \leq 1.0$, which in turn sets a lower limit on $i$. Conversely if 
we set an upper limit on the inclination angle of $i = 90^{\circ}$, we obtain 
a lower limit for $\beta$. Given these two extremes, upper and lower limits on 
the mass of both the neutron star and the optical companion may be calculated, 
as shown in the two left hand columns of Table 1. Solutions lying between the 
two extremes, corresponding to intermediate values of $i$ and $\beta$, are of 
course also valid. 

\subsection{X-ray heating corrections}

RV measurements of the optical companion in a binary system 
reflect its motion about the centre of light. In systems with Keplerian orbits,
the centre of light should be roughly coincident with the centre of 
mass. However, X-ray heating of the optical companion can lead 
to an offset between the two centres, such that the observed RVs may not 
represent the true motion about the centre of mass. In order to 
determine accurate masses from RV curves, these non-Keplerian deviations 
must therefore be accounted for. 

To correct for the heating effects we followed Reynolds et al. 
(1993) and ran models using {\sc light}2 (Hill 1988), a sophisticated 
light-curve synthesis program. In the mode used here, the program generates 
non-Keplerian velocity 
corrections by averaging a velocity based on contributions from elements of 
the giant star's projected stellar disk, where each element is weighted 
according to the flux at that point. We note that the He~I lines will be 
stronger on the cooler side of the star (i.e. away from the X-ray source)
and this may shift the light centre in the opposite direction to the flux
weighting correction which {\sc light2} applies. The net effect of the 
true heating correction may therefore be smaller than calculated here.

Due to the limitations of {\sc light2} we 
were unable to accurately represent the dimensions of the neutron star in the 
model. Instead, we set the radius and polar temperature of the object 
representing the neutron star to produce a blackbody luminosity 
equivalent to the observed X-ray luminosity, which is essentially all that 
matters to calculate the heating correction. We used $L_{\rm x} = 2.4 
\times 10^{38}$~erg~s$^{-1}$  (Paul et al. 2002), which  agrees well with 
the value of $L_{\rm x} = 2.0 \times 10^{38}$~erg~s$^{-1}$ that we determined 
using the {\em RXTE} ASM flux for the epoch of observation, assuming a 
simplified X-ray spectral shape and a distance to the SMC of $D = 60.6$~kpc 
(Hilditch et al. 2005). 

The initial values for $i$ and $q$ input to {\sc light2} were those 
obtained from the Monte Carlo analysis using the raw value for Sk~160's 
RV amplitude. Having calculated the RV correction at the phase 
corresponding to each of the spectra, the individual RV
measurements were adjusted accordingly, and a new solution for the RV 
amplitude was found. This amplitude was fed into the Monte Carlo
program to determine new values for $i$ and $q$, and these values were then 
fed back into {\sc light}2 to recalculate the RV corrections. We found that 
the code was required to run through 3 iterations before there was no 
further change in the heating corrections. The results of this process 
are shown in the  two right hand columns of Table 1 and the final corrected 
RV curve is shown in Figure 3, with the data listed in Table 2.

\subsection{He~II 4686\AA \ emission line}

Hutchings et al. (1977) noted that the He~II 4686\AA \ emission line seen 
from Sk~160 moves approximately in antiphase with respect to 
the He~I lines, and follows the RV of the neutron 
star, albeit with a lower amplitude, i.e. $\sim 250$~km~s$^{-1}$ as opposed 
to $\sim 300$~km~s$^{-1}$. In order to investigate this with our data, we 
performed Gaussian fits to the He~II emission lines in each spectrum, 
and plotted their heliocentric corrected velocities against phase in Figure 4;
the data are listed in Table 2. 
In the more noisy spectra, it was not possible to measure this 
emission line, so there are less data points here than in Figures 2 and 3. 
Overplotted on these data is the best-fit sinusoid which has an amplitude of 
$265 \pm 8$~km~s$^{-1}$, a systemic velocity of $167 \pm 7$~km~s$^{-1}$, and
a phase shift of $0.46 \pm 0.01$ with respect to the ephemeris of Wojdowski 
et al. (1998). What we see is very similar to the behaviour noted by Hutchings 
et al, in that the He~II emission line velocity tracks the motion of the 
neutron star but has a lower amplitude than that of the neutron star itself.
The slight phase shift from the motion of the neutron star is also similar
to that seen in the He~II emission line RV of Cyg X-1 (Gies \& Bolton 1986a, 
1986b). As noted by Hutchings et al, this might indicate an origin for the 
emission that lies between the neutron star and the surface of Sk~160. We 
suggest that a possible site for this emission may be a hot-spot where a 
stream of material accreting via Roche lobe overflow impacts the outer edge 
of the neutron star's accretion disk.

\section{Discussion}

Our final value for $\gamma$ of $173.8 \pm 1.5$~km~s$^{-1}$ is in excellent 
agreement with the value obtained by Reynolds et al. (1993) from their heating
corrected radial velocity curve, namely $173.0 \pm 1.5$~km~s$^{-1}$. However, 
our raw and corrected values found for $K_{\rm o}$ and the corresponding upper 
limits to the neutron star mass (i.e. those corresponding to $\beta=1$) are 
each lower than those found by Reynolds et al. The discrepancy could be due 
to the limited phase coverage of their data set and the fact that the value 
they assume for $L_{\rm x}$, when determining the non-Keplerian corrections, 
is significantly higher than the value we used. In comparison, our raw value 
for $K_{\rm o}$ and the corresponding upper limit on the neutron star mass 
are both in good agreement with those found by van der Meer et al. (2005), 
namely $K_{\rm o} = 20.3 \pm 0.9$~km~s$^{-1}$ and $M_{\rm x} = 1.05 \pm 
0.09$~M$_{\rm \odot}$. In both of these previous studies, however, 
the authors assume that the giant star is Roche-lobe filling. Whilst this is 
plausible, it only gives an upper limit for the mass of the neutron star, as 
noted above.

There remains the question of whether an accretion disk will cast a 
shadow on the giant star so reducing the effect of X-ray heating, as raised 
by van Kerkwijk et al. (1995). In a study of Her X-1, Reynolds et al. (1997) 
corrected for non-Keplerian deviations using both a diskless model 
({\sc light}2) and a disk model. In that case the quantitative agreement 
between the two models was found to be good. As Her X-1 clearly has a disk, 
and also has a companion star that is significantly smaller than that in 
SMC X-1, we can assume that the absence of a disk in the {\sc light}2 code has 
negligible effect on the heating corrections in our case. 

Model calculations of Type II supernovae suggest that these events produce a 
bimodal distribution of initial neutron star masses, with averages 
within those peaks of 1.28 and 1.73~M$_{\odot}$, whereas Type Ib supernovae 
produce neutron stars with masses around 1.32~M$_{\odot}$ (Timmes et 
al 1996). Neutron stars produced in Type Ia supernovae are expected to have 
masses close to the Chandrasekhar limit, 1.44~M$_{\odot}$ (for a He white 
dwarf). Based on our heating-corrected mass determination of 
$(0.91\pm0.08)$~M$_{\odot} - (1.21\pm0.10)$~M$_{\odot}$ (depending on the 
Roche lobe filling factor), the neutron star in SMC X-1 is consistent with 
the first peak in the Type II supernovae bimodal neutron 
star mass distribution. We note that mass determinations which {\em do not} 
account for X-ray heating in SMC X-1 give a small neutron star mass that is 
inconsistent with {\em all} of the predictions above. Note also that these 
theoretical models do not take into account any mass that subsequently 
accretes in a binary system, so the theoretical values for the neutron star 
masses in accreting binaries are even higher.

Finally, we note the implications of our interpretation for the origin 
of the He~II 4686\AA \ emission line. If this arises in a stream-disk impact 
hot spot, it confirms that some of the accretion occurs 
via Roche lobe overflow, as previously surmised. It also suggests a potential 
test of the idea. The accretion disc in SMC X-1 is supposed to precess 
with a period of $\sim 55$~d (e.g. Wojdowski et al. 1998). In this case,
the stream-disk impact site should change its location on this 
period, moving closer to and further away from the neutron star as the 
eccentric disc precesses. Both the equivalent width and the RV 
amplitude of the He~II emission line should therefore vary throughout the 
precession cycle. Unfortunately, our data do not extend over enough
of the super-orbital cycle, nor are they of high enough signal-to-noise to 
test this, but such an investigation would be worth carrying out in future.

\begin{acknowledgements}
We are grateful to the staff of the SAAO for scheduling the observations on 
which this paper is based and their help during the observing run. 
We thank Graham Hill for the use of his {\sc light}2 code and Alastair
Reynolds for the radial velocities from his paper. We are indebted to
Tim Harries for providing us with a linux installation of {\sc light}2 
and Ron Hilditch for his invaluable, patient advice helping us to get
{\sc light}2 to work. We also thank Sean Ryan for assistance with the 
intricacies of cross-correlations in iraf and an anonymous referee for
several helpful suggestions to improve the paper.
\end{acknowledgements}

\clearpage

\setcounter{figure}{0}
\begin{figure*}
\setlength{\unitlength}{1cm}
\begin{picture}(12,10)
\put(3,-3){\includegraphics{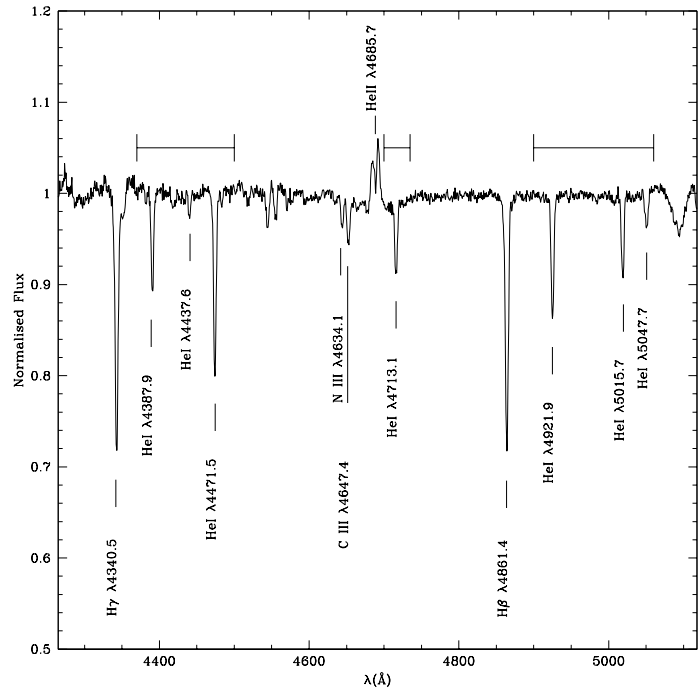}}
\end{picture}
\caption{The median, continuum normalised spectrum of Sk~160. The horizontal 
bars indicate the regions of the spectrum containing He~I absorption lines 
used for cross-correlation and determination of radial velocities.}
\end{figure*}

\setcounter{figure}{1}
\begin{figure*}
\setlength{\unitlength}{1cm}
\begin{picture}(12,10)
\put(-0.5,-5){\includegraphics{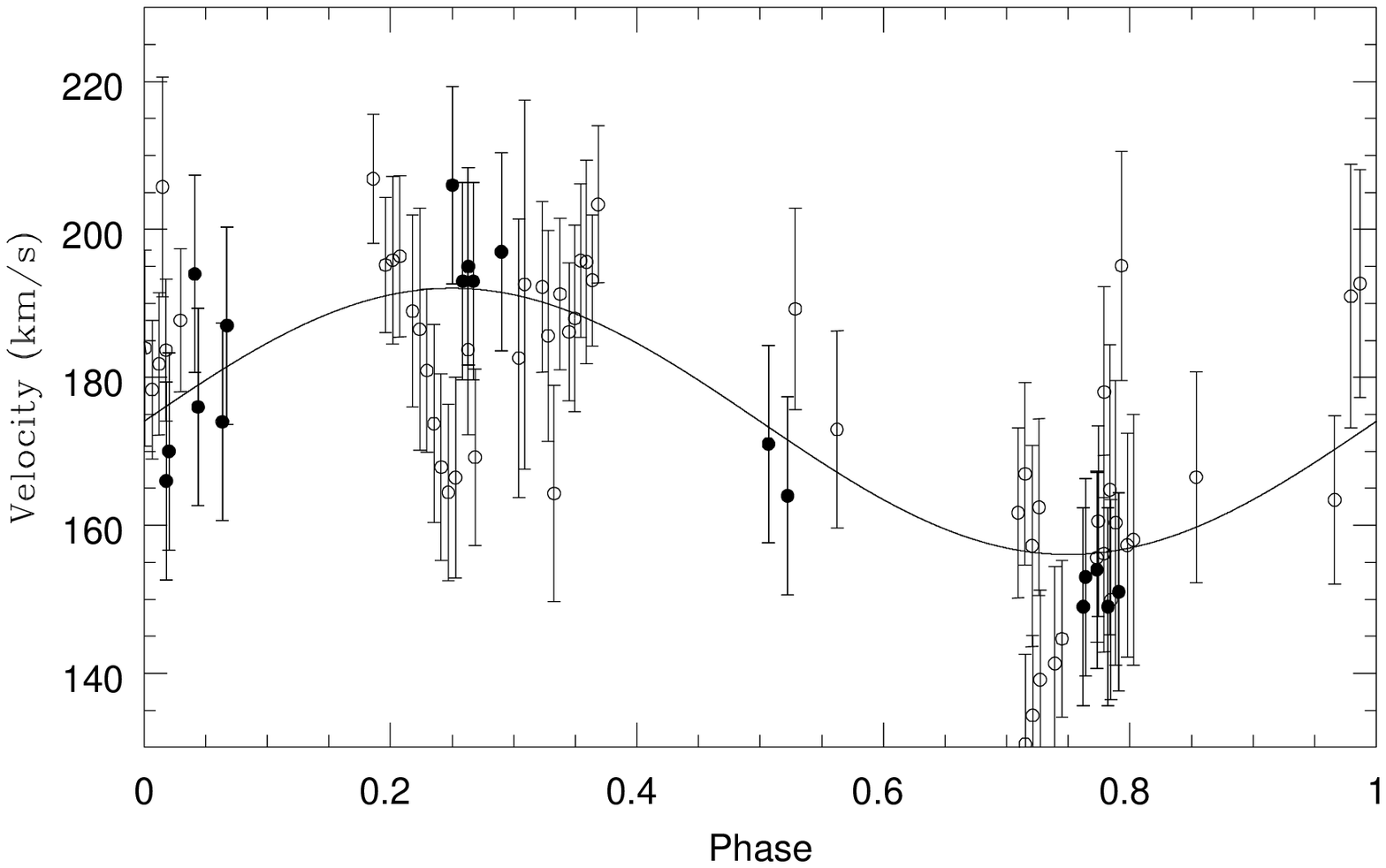}}
\end{picture}
\caption{The uncorrected He~I absorption line RV curve for 
Sk~160. Our data are shown as open circles and those of Reynolds et al. (1993) 
are shown as filled circles. The best fit to the combined data set is 
indicated by the solid line.}
\end{figure*}

\setcounter{figure}{2}
\begin{figure*}
\setlength{\unitlength}{1cm}
\begin{picture}(12,10)
\put(-0.5,-5){\includegraphics{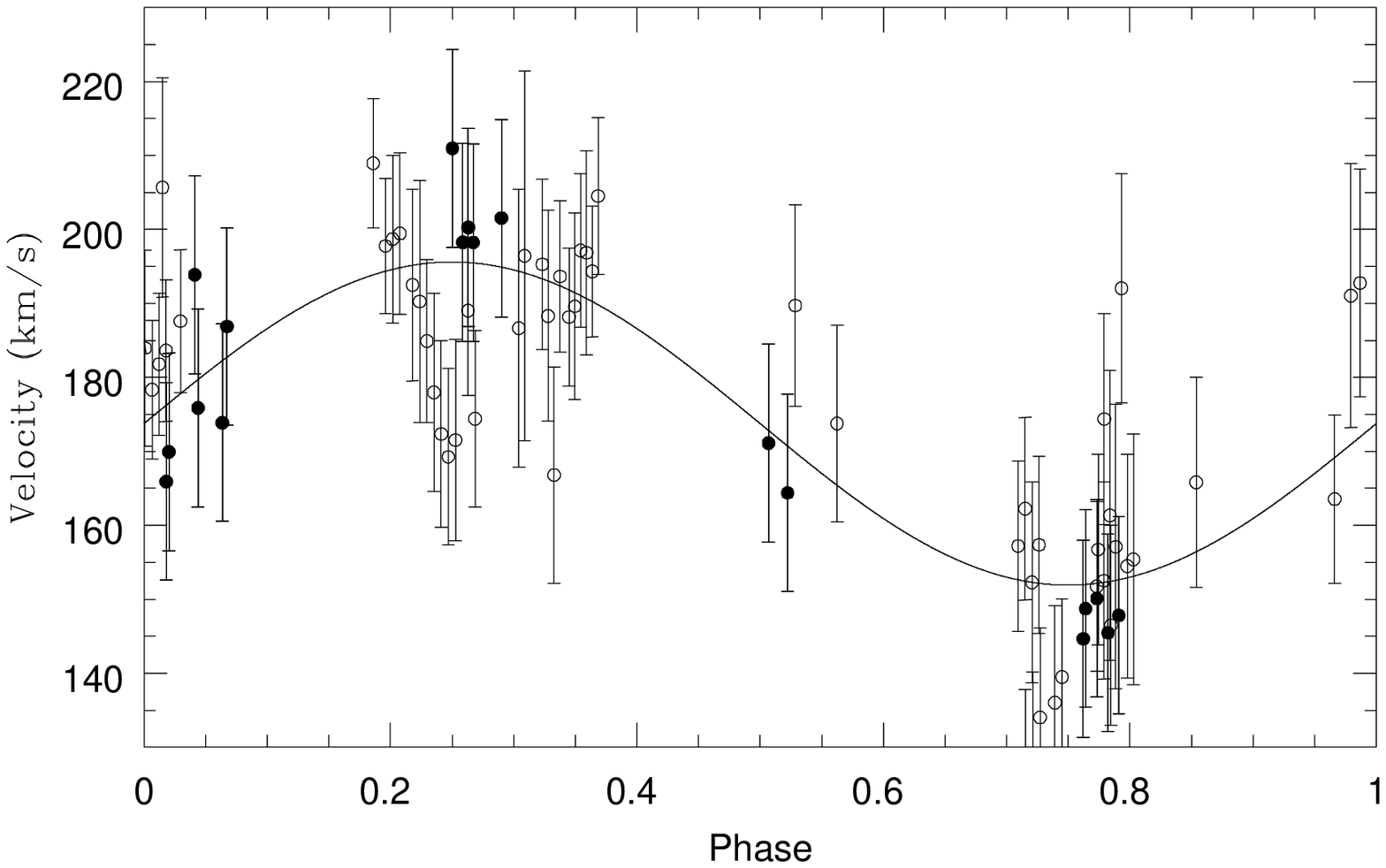}}
\end{picture}
\caption{The heating-corrected He~I absorption line RV curve 
of Sk~160. Our data are shown as open circles and those of Reynolds et al. 
(1993) are shown as filled circles. The best fit to the combined data set 
is indicated by the solid line.}
\end{figure*}

\setcounter{figure}{3}
\begin{figure*}
\setlength{\unitlength}{1cm}
\begin{picture}(12,10)
\put(-0.5,-5){\includegraphics{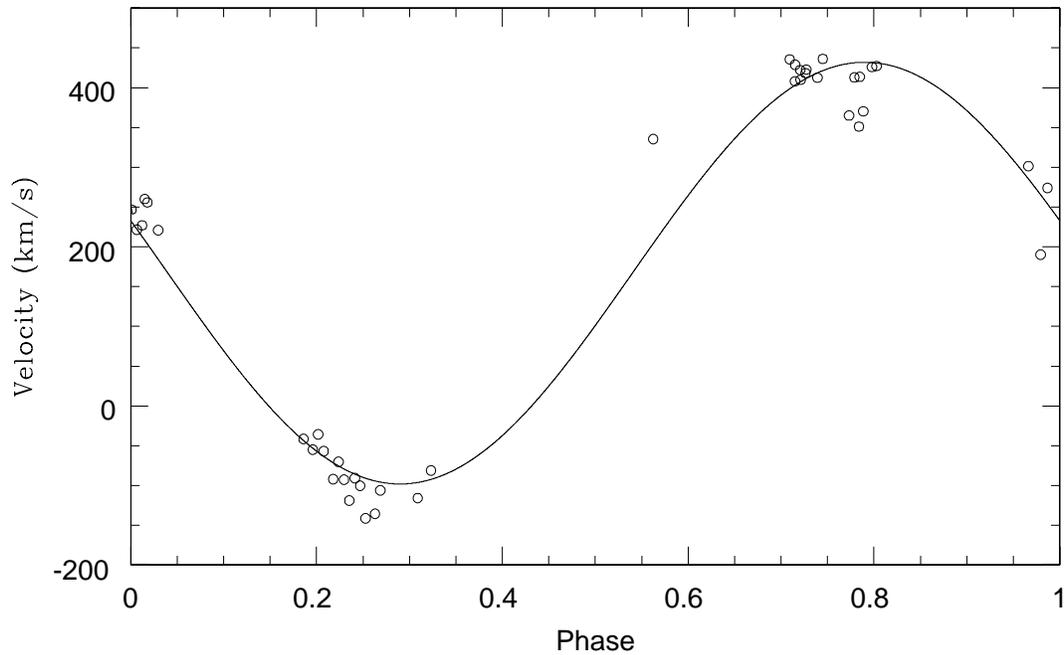}}
\end{picture}
\caption{The RV curve of the He~II 4686\AA \ emission line. The 
solid line indicates a sinusoid with a systemic velocity of 167~km~s$^{-1}$,
an amplitude of 256~km~s$^{-1}$ and a phase shift of 0.46 with respect to
the Wojdowski et al. (1998) ephemeris.}
\end{figure*}

\clearpage

\begin{table*}
\caption{The raw and heating-corrected He~I absorption line heliocentric 
radial velocity data for Sk~160 resulting from the August/September 2000 
observations at SAAO. Also shown are the heliocentric radial velocities 
measured from the He~II 4686\AA \ emision line.}
\begin{tabular}{cccccc} \hline
MJD & Phase & He~I $V_{\rm raw}$ / km~s$^{-1}$ & $1 \sigma$ uncertainty / km~s$^{-1}$ & He~I $V_{\rm corrected}$ / km~s$^{-1}$ & He~II $V$ / km~s$^{-1}$\\ \hline
51786.976 & 0.715 & 130 & 12 & 126 & 429\\
51786.999 & 0.721 & 134 & 11 & 129 & 410\\
51787.022 & 0.727 & 139 & 12 & 134 & 423\\
51787.069 & 0.739 & 141 & 13 & 136 & 413\\
51787.091 & 0.745 & 145 & 11 & 139 & 436\\
51788.808 & 0.186 & 207 &  9 & 209 & --41\\
51788.847 & 0.196 & 195 &  9 & 198 & --55\\
51788.870 & 0.202 & 196 & 11 & 199 & --36\\
51788.892 & 0.208 & 196 & 11 & 199 & --57\\
51788.932 & 0.218 & 189 & 13 & 193 & --92\\
51788.955 & 0.224 & 187 & 16 & 190 & --70\\
51788.977 & 0.230 & 181 & 11 & 185 & --93\\
51789.000 & 0.235 & 174 & 13 & 178 & --119\\ 
51789.022 & 0.241 & 168 & 13 & 172 & --91\\
51789.045 & 0.247 & 164 & 12 & 169 & --100\\
51789.068 & 0.253 & 166 & 14 & 172 & --141\\
51789.108 & 0.263 & 184 & 12 & 189 & --136\\
51789.130 & 0.269 & 169 & 12 & 174 & --106\\
51790.845 & 0.709 & 162 & 12 & 157 & 436\\
51790.867 & 0.715 & 167 & 12 & 162 & 408\\
51790.889 & 0.721 & 157 & 14 & 152 & 422\\
51790.911 & 0.726 & 162 & 12 & 157 & 418\\
51791.094 & 0.773 & 156 & 11 & 152 & 365\\
51791.116 & 0.779 & 156 & 13 & 153 & 413\\
51791.138 & 0.785 & 150 & 14 & 146 & 414\\
51791.845 & 0.966 & 163 & 11 & 164 & 301\\
51791.896 & 0.979 & 191 & 18 & 191 & 190\\
51791.925 & 0.987 & 193 & 15 & 193 & 274\\
51791.979 & 0.001 & 184 & 13 & 184 & 247\\
51792.002 & 0.007 & 178 &  9 & 178 & 222\\ 
51792.024 & 0.012 & 182 & 10 & 182 & 227\\
51792.046 & 0.018 & 184 & 10 & 184 & 256\\
51792.091 & 0.030 & 188 & 10 & 188 & 221\\
51799.819 & 0.015 & 206 & 15 & 206 & 260\\
51801.817 & 0.528 & 189 & 14 & 190 & \\
51802.774 & 0.774 & 161 & 13 & 157 & \\
51802.792 & 0.779 & 178 & 14 & 174 & \\
51802.811 & 0.784 & 165 & 20 & 161 & 351\\ 
51802.829 & 0.789 & 160 & 19 & 157 & 370\\
51802.847 & 0.793 & 195 & 16 & 192 & \\
51802.865 & 0.798 & 157 & 15 & 154 & 426\\
51802.884 & 0.803 & 158 & 17 & 155 & 427\\
51803.083 & 0.854 & 167 & 14 & 166 & \\
51804.836 & 0.304 & 183 & 19 & 187 & \\
51804.854 & 0.309 & 193 & 25 & 196 & --116\\
51804.910 & 0.323 & 192 & 12 & 195 & --81\\
51804.928 & 0.328 & 186 & 14 & 188 & \\
51804.947 & 0.333 & 164 & 15 & 167 & \\
51804.966 & 0.338 & 191 & 10 & 194 & \\
51804.994 & 0.345 & 176 &  9 & 188 & \\
51805.013 & 0.350 & 188 & 13 & 190 & \\
51805.031 & 0.354 & 196 & 10 & 197 & \\
51805.049 & 0.359 & 196 & 14 & 197 & \\
51805.068 & 0.364 & 193 &  9 & 194 & \\
51805.086 & 0.369 & 203 & 11 & 205 & \\
51805.841 & 0.562 & 173 & 13 & 174 & 335\\ \hline
\end{tabular}
\end{table*}

\end{document}